# Mathematical analysis of the spreading of a rumor among different subgroups of spreaders

Raúl Isea[1*] and Rafael Mayo-García[2]

[1]Fundación Instituto de Estudios Avanzados, IDEA. Hoyo de la Puerta, Baruta, Venezuela.

[2]CIEMAT. Avda. Complutense, 40, 28040 Madrid. Spain

(*) Corresponding author: Raúl Isea. Email: risea@idea.gob.ve

**Abstract**

This paper presents a system of differential equations that describes the spreading of a rumor when it is propagated by different subgroups of spreaders. The system that is developed is a generalization of the model proposed by Daley and Kendall. Finally, the system is applied to the exchange rate of the parallel dollar in Venezuela, where the source data that was used were obtained from Google Trends.

**Keywords**: Rumor, Daley and Kendall, parallel dollar, Google Trends.

## Introduction

The following discussion is based on a social action in which we express our point of view. The interest of them is to transmit some information and the need to listen it by others. In fact, many times our communications depend upon the interests of the people who generate this information as well as upon those who are willing to receive it. Usually, we make a decision based on such information in order to verify the validity of it, and at that moment, a rumor could be created. According to the Dictionary of the Royal Spanish Academy (Spanish: Real Academia Española, RAE), a rumor is defined as "*the voice which runs between the public*".

Thanks to Information and Communications Technologies, it is no longer required to have direct conversations between people. It is now possible to hold them through social networks in order to spread the idea, employing, for example, twitter, Facebook, blogs, etc.

On the other hand, the aforementioned act of the dissemination of ideas is very similar to the spreading of an epidemic. Thus, from the initial information, there could be a widespread dissemination of these ideas among different its groups of people who endure different interactions. In this sense, Daley and Kendall established in their published work of 1965, the mathematical basis for modeling the spreading of a rumor and the disease in an epidemic by just emphasizing that similarity [1].

This work proposes a mathematical model to understand the transmission of a rumor among diverse subgroups of spreaders. In this sense, it will explain the behavior of the exchange rate of the parallel dollar in Venezuela under the theory of rumors that have been disseminated in the community. To accomplish this task, the scope of the model proposed by Daley-Kendall [1] will be generalized in order to explain the spreading of a rumor among different groups, *i.e.* rumor spreaders.

**Daley and Kendall's model**

Daley and Kendall proposed in 1965 [1], a mathematical model to simulate the process of the spreading a rumor, the so-called DK model. This model classifies the population into three different groups:

- The ignorant population U which starts a rumor.
- The spreading population V which spreads the rumor.
- The stifler population W. which hears of the rumor and decides not to spread it.

This model, which is depicted in figure 1, assumes that the rumor spreads according to the interaction between the ignorant and the spreader populations with a probability defined by β/N. We defined the degree of acceptance of the rumor with μ. When a spreader interacts with a stifler, the rumor stops spreading and the likelihood of that happening is $\frac{\gamma V(V+W)}{N}$ (see the corresponding details in equations 1-3).

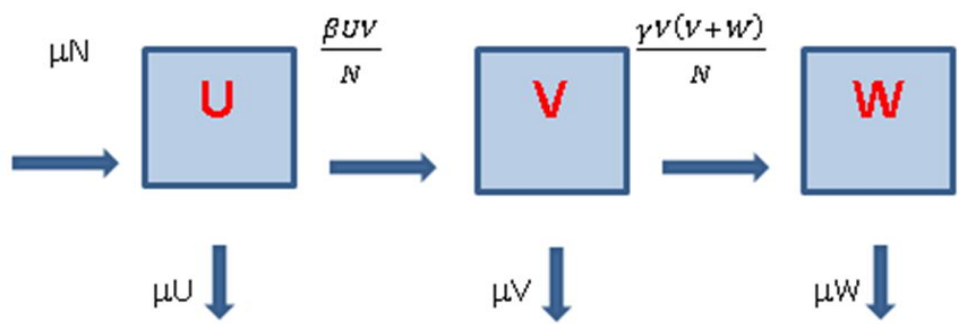

**Figure 1**. Scheme proposed by the Daley and Kendall's model to describe the spreading of rumor (details of the notation are described in the text).

As it might be expected, the rumor loses its value over time [2-5]. Such a probability is defined by the γ factor. This fact is equivalent to considering that the rumor does not remain a novelty, either because it is already known or because it lacks the value to be broadcasted. From all this, the equations describing the model above are given by:

$$\frac{dU}{dt} = \mu N - \frac{\beta UV}{N} - \mu U \qquad (1)$$

$$\frac{dV}{dt} = \frac{\beta UV}{N} - \frac{\gamma V(V+W)}{N} - \mu N \qquad (2)$$

$$\frac{dW}{dt} = \frac{\gamma V(V+W)}{N} - \mu W \qquad (3)$$

The solution of the system of equations (1-3) is:

$$U = \frac{\mu N^2}{(\beta V + \mu N)} \qquad (4)$$

$$V = \frac{N\beta\mu V}{\beta V\gamma + \mu\beta V + \mu^2 N} \qquad (5)$$

$$W = \frac{\gamma \beta^2 N V^2}{(\beta V + \mu N)(\beta V\gamma + \beta\mu V + \mu^2 N)} \qquad (6)$$

Defining $C_1 \equiv \beta V + \mu N$ and $D_1 = \beta V\gamma$, equations (4-6) can be written as:

$$U = \frac{\mu N^2}{C_1} \qquad (7)$$

$$V = \frac{N\beta\mu V}{\beta V\gamma + \mu\beta V + \mu^2 N} = \frac{N\beta\mu V}{\beta V\gamma + \mu C_1} = \left[\frac{\mu N}{\beta V\gamma + \mu C_1}\right]\beta V = \left[\frac{\mu N}{D_1 + \mu C_1}\right]\beta V \qquad (8)$$

$$W = \frac{\gamma \beta^2 N V^2}{(\beta V + \mu N)(\beta V \gamma + \beta \mu V + \mu^2 N)} = \frac{\gamma N \beta^2 V^2}{C_1(D_1 + \mu C_1)}$$

$$= \left[\frac{\gamma N}{C_1(D_1 + \mu C_1)}\right] \beta^2 V^2 \qquad (9)$$

In the next section, the DK model will be generalized to a case where a source of the rumor is spread by subgroups of spreaders.

**The proposed model**

A model to describe the diffusion process of a single rumor (U) between different subgroups of spreaders is proposed. To do that, subpopulations called spreaders are denoted as $V_1$, $V_2$, $V_3$, …$V_n$ (Figure 2). Similar to the DK model, it will also be assumed that each of these spreaders have different probabilities, which are characterized by the speed at which the rumor is transmitted. These rates are defined with the parameters $\beta_1$, $\beta_2$, $\beta_3$, … , $\beta_n$. Finally, the different subgroups of stiflers will be given by $W_1$, $W_2$, $W_3$, …$W_n$. As in the previous case, these stiflers will have a probability of not spreading the rumor, which are defined by the probabilities $\gamma_1$, $\gamma_2$, $\gamma_3$,…, $\gamma_n$ respectively.

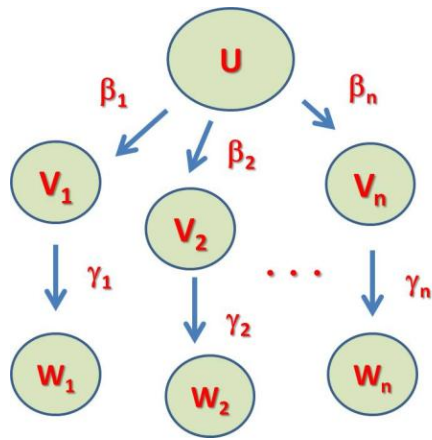

**Figure 2**. Proposed model of the diffusion process of a rumor (U) between *n* different subsets of spreaders, whose probabilities are described by $\beta_n$ and $\gamma_n$ for each spread.

The model proposed in this work is defined with the following differential equations:

$$\frac{dU}{dt} = \mu N - \sum_{n=1}^{T} \frac{\beta_n U V_n}{N} - \mu N \quad (10)$$

$$\frac{dV_n}{dt} = \frac{\beta_n U V_n}{N} - \frac{\gamma_n V_n (V_n + W_n)}{N} - \mu V_n \quad (11)$$

$$\frac{dW_n}{dt} = \frac{\gamma_n V_n (V_n + W_n)}{N} - \mu W_n \quad (12)$$

where n = 1 up to T represents the number of spreaders. In the simplest case described by the DK model of equations (1-3), the value of n and T are equal to 1.

In the present case, different subsets of spreaders are considered (*i.e.* T > 1), so the solution of the system of equations described by (10-12) is:

$$U' = \frac{\mu N^2}{C} \quad (13)$$

$$V'_n = \sum_{n=1}^{T} \left[ \frac{\mu N}{(C_n + \mu C)} \right] \beta_n V_n \quad (14)$$

$$W'_n = \sum_{n=1}^{T} \left[ \frac{\gamma_n N}{C(C_n + \mu C)} \right] \beta_n^2 V_n^2 \quad (15)$$

where the following expression has been taken into account:

$$C \equiv \sum_{n=1}^{T} \beta_n V_n + \mu N$$

$$C_n \equiv \beta_n V_n \gamma_n$$

Thus, if T=2, it means that the rumor is spread between two subgroups of spreaders. The solution of this case is given by:

$$U' = \frac{\mu N^2}{\beta_1 V_1 + \beta_2 V_2 + \mu N} = \frac{\mu N^2}{C}$$

$$V_1' = \left[\frac{\mu N}{C_1 + \mu C}\right] \beta_1 V_1$$

$$V_2' = \left[\frac{\mu N}{C_2 + \mu C}\right] \beta_2 V_2$$

$$W_1' = \left[\frac{\gamma_1 N}{C(C_1 + \mu C)}\right] \beta_1^2 V_1^2$$

$$W_2' = \left[\frac{\gamma_2 N}{C(C_2 + \mu C)}\right] \beta_2^2 V_2^2$$

where $C_1$ and $C_2$ values are $\beta_1 V_1 \gamma_1$ and $\beta_2 V_2 \gamma_2$, respectively. This case may explain how two newspapers handle the same information for example.

**The parallel dollar in Venezuela: a use case**

Finally, the rate exchange of the parallel dollar in Venezuela with this new model is described. To do that, the information obtained from the Internet and employed by Google Trends (http://www.google.com/trends/) with the search term "parallel dollar in Venezuela" is obtained. The reason to use Google Trends is that the Google search engine is one of the major sources of information worldwide, regardless of the veracity of what is posted on it. The results of the query by words are normalised to a [0-100] scale; also, the results can be filtered by geographic area and time. In the presented use case, the time period of study goes from February 2012 to May 2014 and presents only searches performed in Venezuela.

The next step is to consider how many spreaders will be needed to explain the tendency of the parallel dollar exchange rate in Venezuela obtained by Google Trends. In this case, three different subgroups of spreaders (T = 3) are considered, which can be modeled by equations (10-12). In this regard, data from Google Trends correspond to $V_1$, who just spread this information.

The reason for proposing T=3 is trying to reproduce the following scenario, in which there is a first subgroup representing people interested in promoting the illegal market dollar exchange in Venezuela, a second group that respects the exchange regulations for the purchase of foreign currencies, and finally, a group that represents the Government of Venezuela, who seeks to stabilize and normalize the equitable exchange trend in the country.

The next step is to solve this seven-differential equation system according to (10-12). In addition, the values of $\beta$ and $\gamma$ in each of the equations were least squares fitted using the shooting method already described in [6,7] is shown in figure 3, where red circles represent the results obtained from the system of differential equations and the blue solid line is a least squares fitting to the latter in order to find their evolution in time, while the data obtained for Google Trends are represent by green circles. The final result clearly indicates how the trend of the exchange rate of the parallel dollar in Venezuela follows the common rumor propagation model.

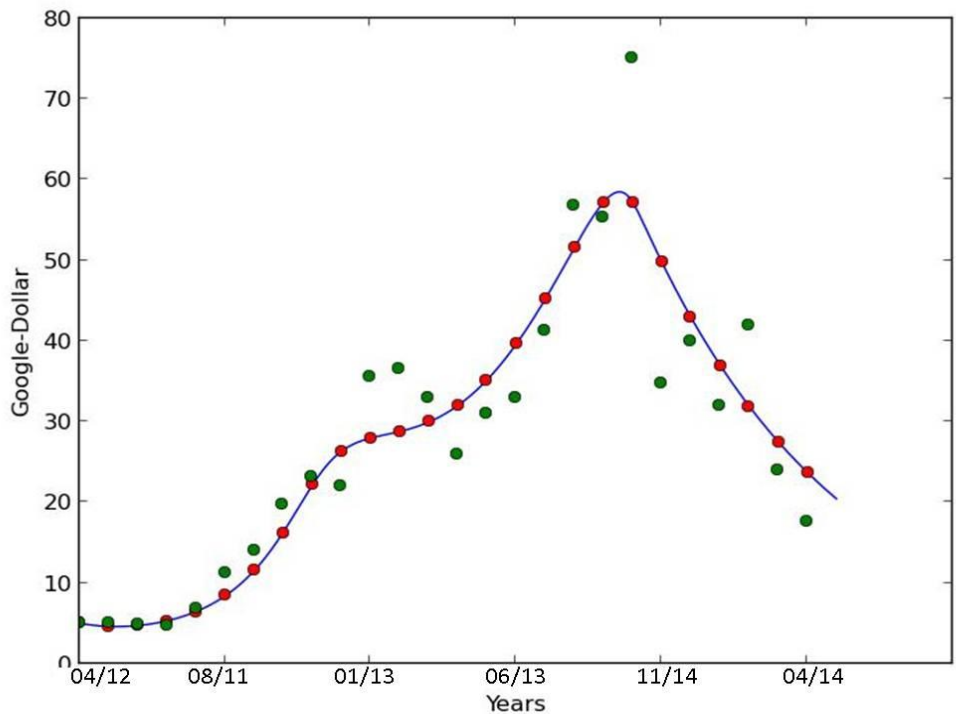

**Figure 3**. Least-squares fitting process of the system of equations that models the rumor concerning parallel dollar in Venezuela from 02/12 to 04/14, considering three different subgroups, *i.e.* T=3.

**Conclusions**

The Information and Communications Technologies are allowing rumors to rapidly spread regardless of the source of the information. Therefore, it is necessary to develop mathematical models to analyze and predict the spreading of the rumors as a function of time. In this regard, it has resulted in a very useful modification of the original scheme proposed by DK.

By applying a mathematical model, it has been possible to explain the diffusion process among various groups of the population from an initial rumor. Thus, the evolution of the news published in Internet in Venezuela about the parallel dollar clearly indicates that it behaves like a rumor. In this point, it is worth mentioning to indicate that it cannot be concluded that all the information obtained from Google Trends is always a rumor, since it is unknown the criterion that Google uses to obtain this value, and *per se* it is a rumor for our consideration. Anyway, this behavior of the parallel dollar has been modeled by using three groups of spreaders. As future work, it will be of interest to develop a new methodology to better assess the quality of the data fitting process. Also, it will be useful to foresee which is the optimum number of spreaders to properly describe how the exchange rate value behaves (*i.e*. the T optimal value). Finally, a study of the stability of this model is warranted.

**Acknowledgement**

The author wishes to express his sincere thanks to Prof. Karl E. Lonngren for their unconditional help and the comments concerning the manuscript. This paper is dedicated to the memory of Raimundo Villegas (founder of the Fundación Instituto de Estudios Avanzados - IDEA) who died on 21th October 2014.

**References**

[1] Daley DJ, Kendall DG, Stochastic rumors, J. Inst. Math. Appl. 142(1965), pp. 42-55.

[2] Piqueira J, Rumor propagation model: An equilibrium study, Math. Probl. Eng. 2010 (2010), ID 631357.

[3] Huo L, Huang P, Guo C, Analyzing the Dynamics of a Rumor Transmission Model with Incubation, Discrete Dyn. Nat. Soc. 2012(2012), ID 328151.

[4] Fedewa N, Krause E, Sisson A, Angelos J, Spread of a rumor. SIAM Undergraduate Research Online (SIURO) 2013 [access jun 1, 2014]. Available at http://goo.gl/ypAe9B

[5] Dietz K, Epidemics and rumors: A survey, J. R. Stat. Soc.  A 130(1967), pp. 505-528.

[6] Wu H, Xue H, Kumar A, Numerical discretization-based estimation methods for ordinary differential equation models via penalized spline smoothing with applications in biomedical research. Biometrics, 68(2012), pp. 344-352.

[7]  Kiusalaas J, Numerical Methods in Engineering with Python. 2nd ed. Cambridge University Press; New York, NY, USA: 2010.